\documentclass[showpacs,prl,twocolumn]{revtex4}
\usepackage{graphicx}
\usepackage{enumerate}
\usepackage{hyperref}
\usepackage{textcomp}
\usepackage{amsmath}
\usepackage{amssymb}
\usepackage{pgf}
\usepackage{diagbox}
\usepackage{booktabs}

\def\bra#1{\langle#1 |}
\def\ket#1{| #1\rangle}

\def\ud{\mathrm{d}}

\newcommand{\nep}{\textrm{e}}

\newcommand{\QA}{\mathrm{\scriptscriptstyle QA}}

\newcommand{\bfr}{{\bf r}}
\newcommand{\x}{{\bf x}}
\renewcommand{\d}{{\bf d}}
\newcommand{\y}{{\bf y}}
\renewcommand{\l}{{\bf l}}

\newcommand{\K}{{\bf K}}

\newcommand{\A}{{\bf A}}
\newcommand{\calA}{{\mathcal{A}}}
\newcommand{\calB}{{\mathcal{B}}}
\newcommand{\calH}{{\mathcal{H}}}
\newcommand{\calC}{{\mathcal{C}}}
\newcommand{\calP}{{\mathcal{P}}}

\newcommand{\opcdag}[1]{{\hat{c}^{\dagger}}_{#1}}
\newcommand{\opc}[1]{{\hat{c}^{\phantom \dagger}}_{#1}}

\begin{document}

\title{Quantum annealing and non-equilibrium dynamics of Floquet Chern insulators}

\author{Lorenzo Privitera$^{1}$, Giuseppe E. Santoro$^{1,2,3}$}

\affiliation{
$^1$ SISSA, Via Bonomea 265, I-34136 Trieste, Italy\\
%\item 
$^2$ CNR-IOM Democritos National Simulation Center, Via Bonomea 265, I-34136 Trieste, Italy\\
%\item 
$^3$ International Centre for Theoretical Physics (ICTP), P.O.Box 586, I-34014 Trieste, Italy
}

\begin{abstract}
Inducing topological transitions by a time-periodic perturbation offers a
route to controlling the properties of materials. 
Here we show that the adiabatic preparation of a non-trivial state involves a selective population of edge-states, 
due to exponentially-small gaps preventing adiabaticity.
We illustrate this by studying graphene-like ribbons with hopping's phases of slowly increasing amplitude, 
as, e.g., for a circularly polarized laser slowly turned-on.
The induced currents have large periodic oscillations, but flow solely at the edges upon time-averaging, 
and can be controlled by focusing the laser on either edge.
The bulk undergoes a non-equilibrium topological transition, as signaled by 
a local Hall conductivity, the Chern marker introduced by Bianco \& Resta in equilibrium. 
The breakdown of this adiabatic picture in presence of intra-band resonances is discussed.
\end{abstract}
%      El.Str.Graphene   QHE      Topo      QPT
\maketitle

%---------------------------------
%\section{Introduction} \label{sec:intro}
%---------------------------------
{\em Introduction.} Recent experiments \cite{Jotzu_Nature14} have mapped the phase diagram of the Haldane 
model \cite{Haldane_PRL88}, a prototypical Chern insulator, by driving ultra-cold fermionic 
atoms in an optical honeycomb lattice periodically modulated in time.  
Since the early suggestion by Oka \& Aoki \cite{oka2009photovoltaic} for a photovoltaic
Hall effect in graphene, theoretical research aimed at studying topological transitions induced by an 
external periodic perturbation 
--- the so-called {\em Floquet topological insulators} \cite{Cayssol_Pss13} --- 
has been intense \cite{Kitagawa_PRB10,calvo2011tuning,Kitagawa_PRB11,Galitski_NatPhys11,Rudner_PRX13,
Kundu_PRL14,Gomez_PRB14,Torres_PRL14,Grushin_PRL14,Dalessio_NatCom15,
Dalhaus_PRL15,Sentef_Naturecomm15,Titum_PRL15,Caio_PRL15,Dehghani_PRB15}.
Preparing a topologically non-trivial Floquet insulator, out of a standard band insulator, requires
passing through a phase transition point, where the bulk energy gap momentarily closes 
and edge-states start crossing it.
It is usually assumed that this can be done by keeping the system arbitrarily close to
its Floquet ``ground state'' (GS), a concept which we will clarify later on, 
provided the strength of the periodic perturbation is ramped-up 
in an adiabatic way \cite{Breuer_PLA89,Breuer_ZPD89,Kitagawa_PRB11} 
--- realizing a generalized form of quantum annealing (QA) \cite{Kadowaki_PRE98,Santoro_SCI02,Santoro_JPA06,Dutta:book}.
%until the topological transition point is crossed.
%
In this paper we study the QA dynamics of the Haldane model across its topological transition, 
and that of periodically driven graphene-like ribbons, e.g., irradiated by a circularly polarized laser. 
We find that the topological transition comes with an ingredient that makes it different from the 
Kibble-Zurek (KZ) paradigm \cite{kibble80,zurek85} describing the crossing of ordinary critical points \cite{zurek96,Polkovnikov_RMP11,Dutta:book}: 
an exponentially small Landau-Zener (LZ) \cite{Landau_PZS32_1,Zener_PRS32} avoided-crossing gap 
between edge states, which forbids edge-state electrons from adiabatically following the GS, 
no matter how slowly the critical point is crossed.
%
%A similar situation is encountered by QA in quantum spin glasses with random first-order transition, also affected 
%by exponentially small gaps \cite{Zamponi_QA:review}.
%
The peculiarity of this QA dynamics is reflected in non-equilibrium currents flowing at the edges, 
%the initial population of edge states is simply freezed into the system, 
%the states just acquiring a dispersion after the transition, hence becoming current-carrying edge states.
%
%in the laser-irradiated model, they can also 
which could be controlled, e.g. by a laser focusing on the edges.
We also show how the change in the topology of the non-equilibrium state is effectively signaled by  
the dynamical counterpart of a local Chern marker, introduced by Bianco \& Resta \cite{Bianco_PRB11} 
as an indicator of an equilibrium non-trivial bulk topology.
%while  a ``no-go theorem'' \cite{Dalessio_arxiv14} suggests that the usual first Chern integral is not a suitable indicator.

{\em Model and idea.} Graphene-like systems display remarkable properties associated to the pseudo-spin-1/2 $\calA-\calB$
sublattice degree of freedom of the honeycomb lattice, with relativistic Dirac cones
sitting at the two corners  ${\K}_{\pm}=(\frac{2\pi}{\sqrt{3}a},\pm \frac{2\pi}{3a})$ 
--- $a$ being the lattice constant --- of the hexagonal Brillouin Zone (BZ), 
when inversion symmetry (IS) and time-reversal (TRS) are unbroken. 
A minimal single-orbital tight-binding model, allowing for a time-dependence 
in the nearest-neighbor (nn) hopping phase and in the on-site energy, is given by the following Hamiltonian
(omitting spin indices):
\begin{equation} \label{H_gen:eqn}
\hat{H}(t) = t_1 \sum_{(ij)} \nep^{-i \Phi_{ij}(t)} \opcdag{j} \opc{i} + 
\Delta_{AB}(t) \sum_{i} (-1)^i \opcdag{i} \opc{i} \;,
\end{equation}
where $\opcdag{i}$ creates a particle at site $i$, $(ij)$ denotes sums over nn 
and $(-1)^i=+1/\!\!-\!1$ on $\calA/\calB$.
$\Delta_{AB}$ controls IS, opens up a trivial equilibrium gap at the Dirac points and is in principle
controllable versus time in optical lattice experiments \cite{Jotzu_Nature14}.
The phases $\Phi_{ij}(t)$ --- generally breaking TRS --- may
result from a time-periodic modulation of the optical lattice in the neutral cold atoms experiments \cite{Jotzu_Nature14},
or from the Peierls' substitution minimal coupling of the electrons 
with the (classical) electromagnetic field of a laser
%in irradiated electronic systems  
$\Phi_{ij}(t)= \frac{e}{\hbar c} \int_{i}^{j} \! d\l \cdot \A(\x,t)$. 
%$t_1\sim -2.8$ eV being the n.n. hopping. 
With the latter realization in mind, we take the field monochromatic and described by a vector potential 
$\A(\x,t)=A_0(\x,t) \left[\hat{\x}\sin(\omega t) + \hat{\y}\sin(\omega t-\varphi)\right]$,
where $\varphi$ describes a general elliptical polarization of the laser %, allowing for
%linearly ($\varphi=0,\pi$, at $\pm 45^{\circ}$) and circularly polarized ($\varphi=\pm \pi/2$) beams.
%
and $A_0(\x,t)$ is a smooth function of space and time.
 %, allowing for a spatially focused beam which is slowly turned on. 

\begin{figure}
\begin{center}
\includegraphics[width=8.5cm]{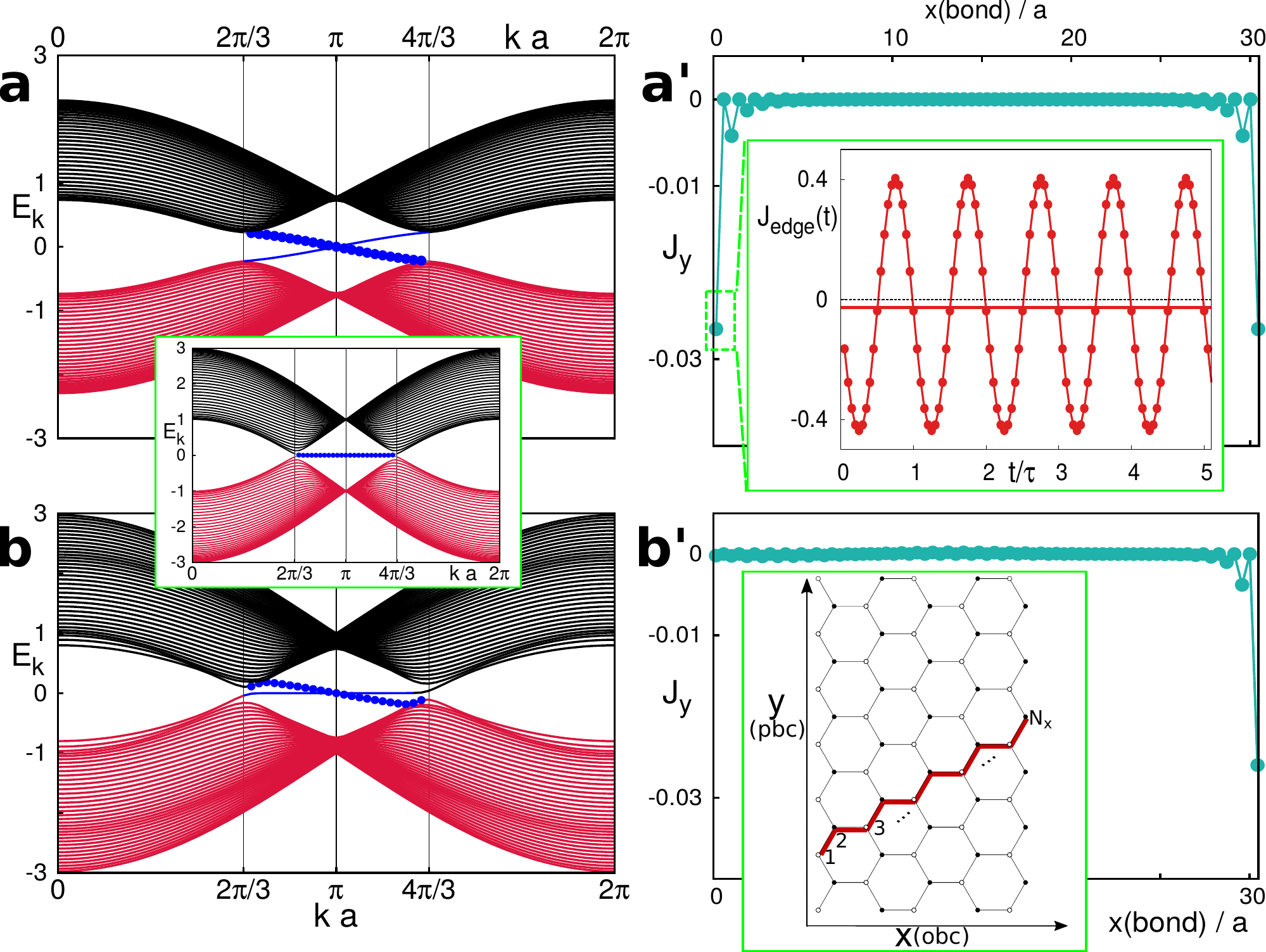}
\end{center}
\setlength{\abovecaptionskip}{0cm}
\caption{(Color online) 
(${\bf a}$): Final Floquet quasi-energy bands for a uniform driving with $\phi = - \pi/2$, $\hbar \omega = 7 |t_1| > W$,
$\Delta_{AB}=10^{-3}|t_1|$ (effectively representing graphene), 
and $\lambda(t)$ linearly ramped up to $\lambda_{\rm f}=1$ in $\tau_{\QA}=100 \tau$, with $\tau=2\pi/\omega$.
Here the topological transition occurs at $\lambda_{\rm cr}\approx 0$. 
Valence states (in red) are filled, conduction states (in black) empty.
Filled circles denote occupied edge states.
Inset: the initial graphene spectrum at $\lambda(0)=0$. 
(${\bf a'}$): Time-averaged bond currents calculated with a periodic evolution at constant $
\lambda_{\rm f}$ for $\tau_{\rm f}= 100 \tau$. 
The inset shows the large intra-period oscillations.
(${\bf b}, {\bf b'}$): Same as ${\bf a}, {\bf a'}$ for an inhomogeneous driving focused on the right edge ($x_c=L_x$) of width $\sigma=0.4L_x$. In the inset, a sketch of the zig-zag strip.
}
\label{Graphene1:fig}
\end{figure}

For a circularly polarized ($\varphi=\pm \pi/2$) spatially uniform laser  \cite{oka2009photovoltaic,Kitagawa_PRB11,Dehghani_PRB15,Sentef_Naturecomm15},
$\Phi_{ij}(t)=\lambda(t) \sin(\omega t +\phi_{ij})$, with $\phi_{ij}=(\pm \frac{\pi}{3}, \mp \frac{\pi}{3}, \pi)$ 
along the 3 nn directions $(\d_1,\d_2,\d_3)$
%$\d_1=d(1/2,-\sqrt{3}/2)$, $\d_2=d(1/2,\sqrt{3}/2)$ and $\d_3=d(-1,0)$ 
connecting an $\calA$-site to its nn $\calB$-sites, 
and $\lambda(t)=\frac{ed}{\hbar c} A_0(t)$, $d$ being the nn distance.
%(For a uniform circularly polarised laser, $\lambda(t)=\frac{ed}{\hbar c} A_0(t)$, $d$ being the nn distance.)
If the frequency $\omega$ is larger than the unperturbed bandwidth $W=6|t_1|$, and 
$\lambda(t)$ and $\Delta_{AB}(t)$ are nearly constant during a period $\tau=2\pi/\omega$, 
the resulting Floquet evolution operator $\hat{U}(\tau,0)=\nep^{-i\hat{\calH}^F\!\tau/\hbar}$ 
has an effective Floquet Hamiltonian $\hat{\calH}^F$ approximately given by a 
Haldane model $\hat{H}_H$ with flux $\phi_H=\pm \frac{\pi}{2}$, 
the same on-site difference $\Delta_{AB}$, and hoppings renormalized by Bessel functions:
$t_{1} \to t_1 J_0(\lambda)$, $t_{2}=-\sqrt{3} [t_1 J_1(\lambda)]^2/(\hbar\omega)$ \cite{Kitagawa_PRB11}.
As the amplitude $\lambda(t)$ is slowly turned-on %of the phase modulation
--- and/or $\Delta_{AB}(t)$ is slowly decreased to $0$ --- 
we effectively drive the Haldane model $\hat{H}_H(t)$ across its equilibrium critical point 
$(\Delta_{AB}/t_2)_c=3\sqrt{3}$ \cite{Haldane_PRL88}. 
%separating the trivial from non-trivial insulating state in its equilibrium phase diagram \cite{Haldane_PRL88}. 
%
%This is, in essence, the setting of a quantum annealing (QA) dynamics,  
%%originally introduced \cite{Finnila_CPL94} as an alternative to classical simulated annealing 
%%\cite{Kirkpatrick_SCI83} for finding the minimal energy state of complex multidimensional problems
%\cite{Kadowaki_PRE98,Santoro_SCI02,Santoro_JPA06,Dutta:book}.
%%closely related to the so-called Adiabatic Quantum Computation \cite{Farhi_SCI01}.
In what follows, we study zig-zag strips with open boundary conditions (OBC) and $N_x$ sites  
in the $x$-direction, and periodic BC (PBC) along $y$ (see inset in Fig.~\ref{Graphene1:fig}-${\bf b'}$). 
For each of the $N_y$ $y$-momenta $k$, the single-particle Hamiltonian
is a $N_x\times N_x$ matrix ${\mathbb H}(k,t)$
%
%Introducing Bloch transformations
%$\opcdag{i,k}=\frac{1}{\sqrt{N_y}}\sum_{ j}^{N_y} \nep^{ikaj} \opcdag{i,j}$, we express the Hamiltonian as
%$\hat{H}(t) = \sum_{k}^{{\rm BZ}_y} \hat{H}(k,t)^{\phantom \dagger}$. 
%
whose Schr\"{o}dinger dynamics is numerically integrated with a $4^{th}$-order Runge-Kutta method,  
the initial state $\ket{\Psi(0)}$ being the Slater determinant GS of $\hat{H}(0)$ at half filling
\cite{SM:note}. 
%(since $\hat{H}(t)$ is quadratic, $\ket{\Psi(t)}$ will be a SD at all $t$),
%see Supplementary Material (SM) for details.

\begin{figure}
\begin{center}
  \includegraphics[width=8.5cm]{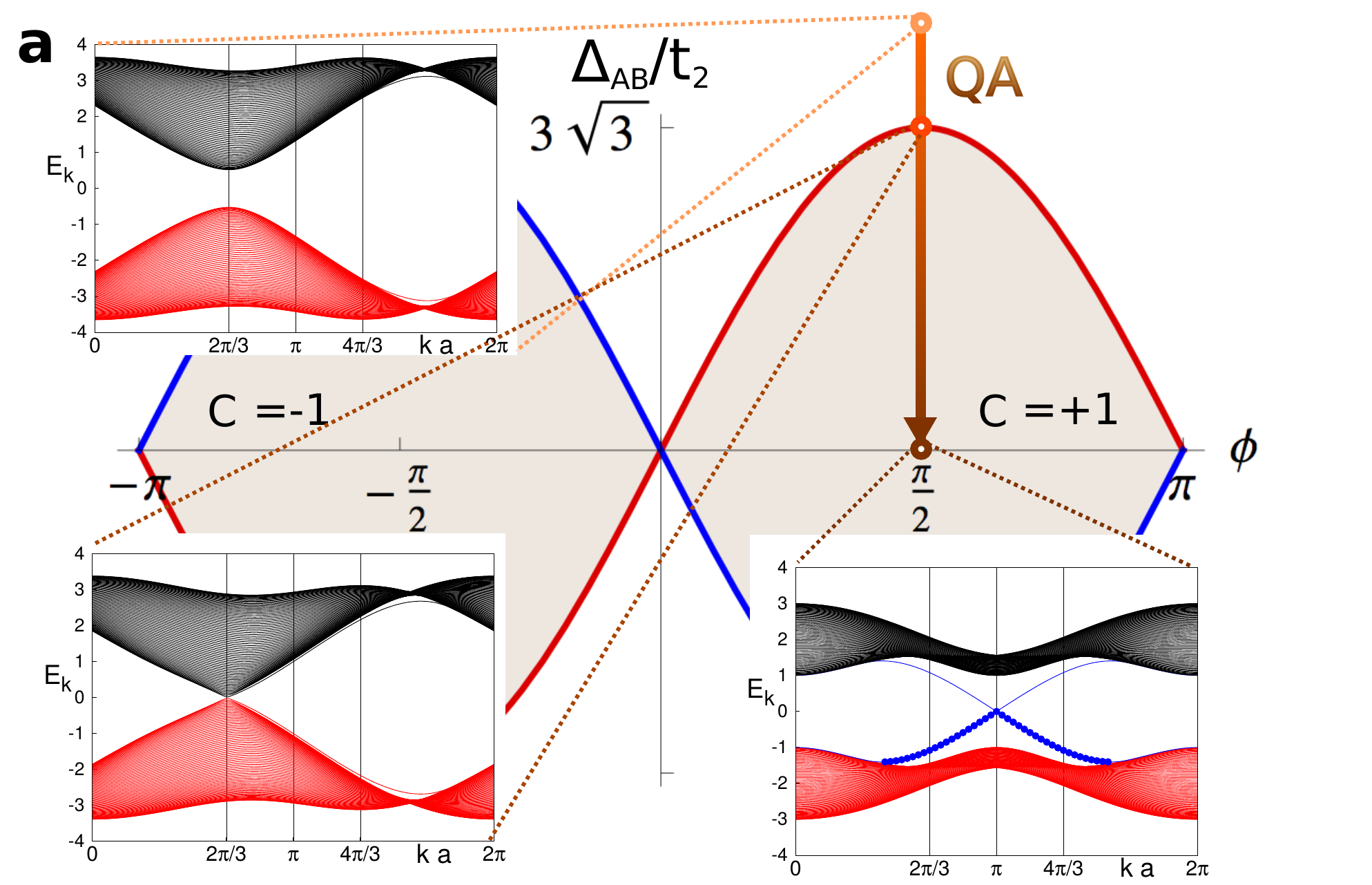}\\ 
  \includegraphics[width=8.5cm]{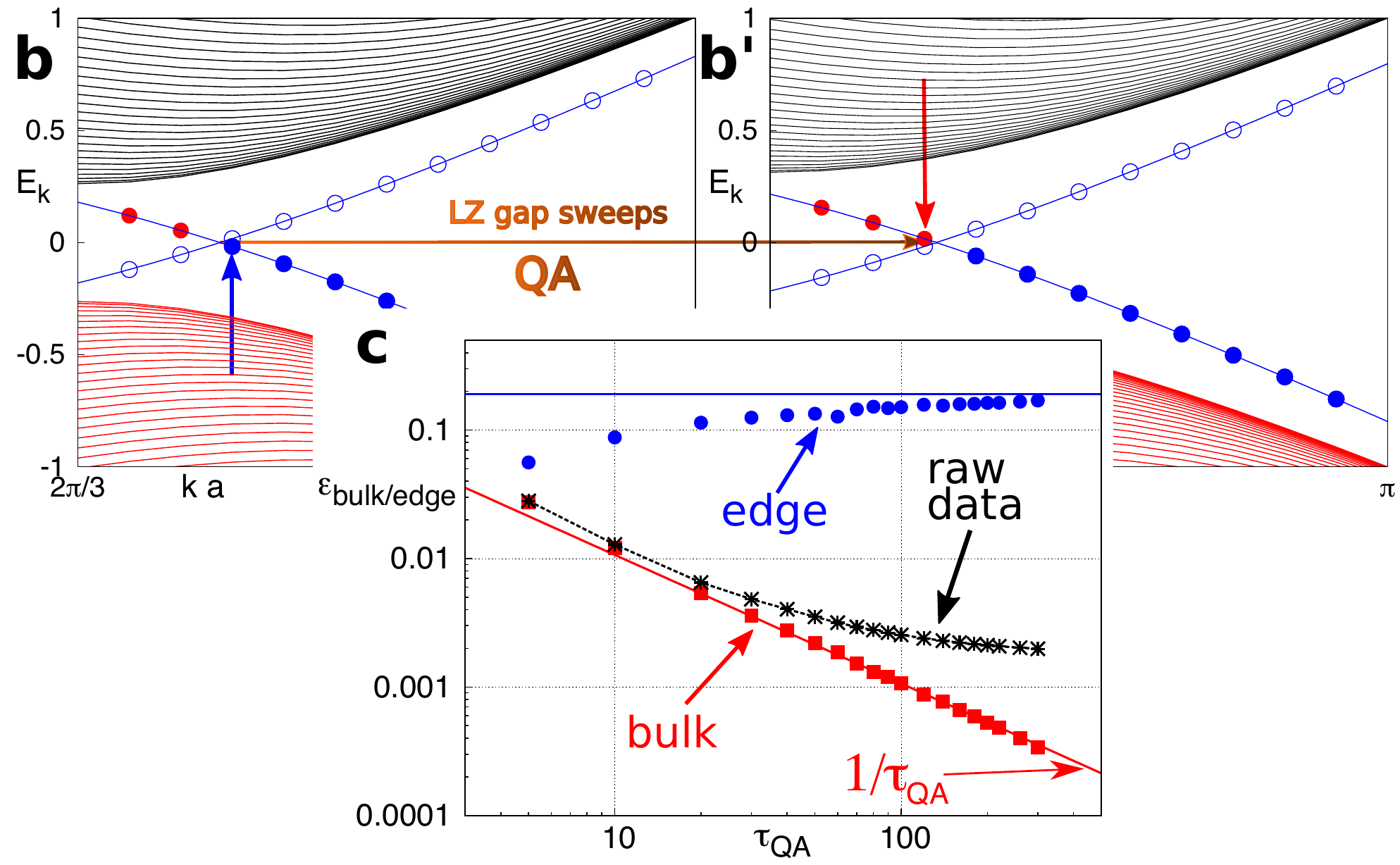}
\end{center}
\setlength{\abovecaptionskip}{0cm}
\caption{(Color online)
(${\bf a}$): Phase diagram of the Haldane model, with three representative
zig-zag strip spectra, along the path of the QA evolution (arrow).
(${\bf b}$, ${\bf b'}$): The mechanism by which right-edge states get selectively populated 
as the exponentially-small LZ gap sweeps to larger values of $k$ during the QA evolution.
The (equilibrium) edge states shown refer to $\Delta_{AB}/t_2=2.5$ (for ${\bf b}$) and 
$\Delta_{AB}/t_2=2.4$ (for ${\bf b'}$).
Filled/empty circles denote occupied/empty edge states.
The blue vertical arrow in ${\bf b}$ points to an occupied right-edge electron that remains 
occupied (red vertical arrow in ${\bf b'}$) after the Landau-Zener event.
(${\bf c}$): Residual energy (stars), separated into bulk (squares) and edge (circles) contributions, vs the annealing time $\tau_{\QA}$. 
Here data with $L=N_x=N_y =6n$ from $18$ to $102$ were used to get $\varepsilon_{\rm bulk/edge}$ for each $\tau_{\QA}$.}
\label{FigHaldane:fig}
\end{figure}

As a warm-up, let us first consider a QA of the Haldane model. %, which illustrates in a physically
%transparent way the mechanism behind the selective non-equilibrium excitation of edge states we have discovered.
% 
%The present findings not only will set the stage for more complicated cases considered next, but
%are also directly amenable to experimental tests with the techniques of Ref.~\cite{Jotzu_Nature14}.
% 
Its phase diagram \cite{Haldane_PRL88}, $\Delta_{AB}/t_2$ vs $\phi_H$, is shown in Fig.~\ref{FigHaldane:fig} {\bf a},
the shaded regions denoting the topologically non-trivial phases with Chern number $\calC=\pm 1$.
In the insets, we show three zig-zag spectra at $\phi_H=\frac{\pi}{2}$:
a trivial insulator with $\Delta_{AB}/t_2=4\sqrt{3}$, the critical point $(\Delta_{AB}/t_2)_c=3\sqrt{3}$, and 
the IS-symmetric point with $\Delta_{AB}=0$. 
%\footnote{(Assuming the strip width $N_x$ to have an even number of sites, we take the two zig-zag edges to start with
%$\calA$-sites on the left, and end with $\calB$-sites on the right. $\Delta_{AB}\neq 0$ breaks
%the left-right symmetry.)}.
%filled/empty circles represent occupied/unoccupied edge states. 
Edge states cross the bulk gap in the non-trivial phase; 
%for $\phi_H=\frac{\pi}{2}$, left (right) edge states have a positive (negative) $k$-dispersion, 
%hence lead to an equilibrium positive (negative) particle current flowing at the left (right) edge.
%
the crossing $k$-point between the two branches moves from the bulk-projected Dirac point $K_+=2\pi/(3a)$ 
towards $K_{\rm f}=\pi/a$ as $\Delta_{AB}/t_2$ decreases from $(\Delta_{AB}/t_2)_c$ to $\Delta_{AB}=0$.
Actually, this is an {\em avoided-crossing} LZ point, with an {\em exponentially small} gap 
$\sim e^{-L_x/\xi}$, where $L_x$ is the strip width and $\xi$ the localization length of the edge states, 
separating the two quasi-degenerate edge states.
Consider the evolution denoted by the arrow in Fig.~\ref{FigHaldane:fig}-{\bf a}: 
$\Delta_{AB}/t_2$ starts from $4\sqrt{3}$ and ends in $0$ in a time $\tau_{\QA}$.
In the initial part of the evolution, the Dirac-point (bulk) gap $\Delta_{K_{\pm}}$ closes at criticality as 
$\Delta_{K_{\pm}} \sim 1/L$, resulting in a standard KZ \cite{kibble80,zurek85} 
non-adiabatic excitation of electrons into the conduction band \cite{Dutta:book}.
In 2d, the critical exponents $\nu=1$, $z=1$ should lead to a residual energy 
$E_{\rm res}(t) =\langle \Psi(t)|\hat{H}(t)|\Psi(t)\rangle-E_{\rm gs}(t)$,
%
%\begin{equation} \label{E_res:eqn}
%E_{\rm res}(\tau_{\QA}) =\Big[ \langle \Psi(t)|\hat{H}(t)|\Psi(t)\rangle-E_{\rm gs}(t)\Big]_{t=\tau_{\QA}} \;, 
%\end{equation}
%
%$|\Psi(t)\rangle$ denotes the state evolved according to the time-dependent Schr\"odinger equation, 
%and
$E_{\rm gs}(t)$ being the instantaneous GS energy,
scaling as $\varepsilon_{\rm res}=E_{\rm res}(t=\tau_{\QA})/L^2 \sim \tau_{\QA}^{-1}$.
The bare data for $\varepsilon_{\rm res}$ (stars in Fig.~\ref{FigHaldane:fig}-{\bf c}) 
depart from this KZ scaling, due to a mechanism of selective edge-state excitation which we now discuss.
%
%The reason for the disagreement sheds light onto a universal mechanism of edge-state selective excitation which we now discuss. 
%As previously mentioned, the avoided-crossing point ``sweeps to the right'', 
%towards $K_{\rm f}=\pi/a$, for the QA evolution we are considering. 
%
Consider the right-edge electron sitting immediately to the right of the LZ gap in Fig.~\ref{FigHaldane:fig} ${\bf b}$
(here we have only $N_y=72$ $k$-points for clarity of illustration):
as the LZ gap sweeps towards larger $k$, it will be {\em unable to follow the ground state} 
due to the exponentially-small LZ gap, and will remain in the right-edge band,
see Fig.~\ref{FigHaldane:fig}-${\bf b'}$, but {\em excited}, since the equilibrium lowest-energy 
state sits in the left-edge band. 
In essence: {\em there cannot be any LZ tunneling across the opposite edges of the sample}. 
Hence, left-edge states remain, one after the other, selectively unoccupied. 
%unlike their fellows on the right-edge.  
If we separate the contributions due to bulk and edges, $E_{\rm res}=\varepsilon_{\rm bulk}L^2 + \varepsilon_{\rm edge}L$,
%--- which scales as $\varepsilon_{\rm edge} L$, as opposed to the bulk $\varepsilon_{\rm bulk} L^2$ --- 
%and analise the QA data as $E_{\rm res}=\varepsilon_{\rm bulk}L^2 + \varepsilon_{\rm edge}L+\cdots$,
we find that $\varepsilon_{\rm bulk}\sim \tau_{\QA}^{-1}$, filled squares in Fig.~\ref{FigHaldane:fig}-${\bf c}$, 
while $\varepsilon_{\rm edge}$ (solid circles) slowly increases, approaching the value 
%(horizontal line in Fig.~\ref{FigHaldane:fig}-${\bf c}$)
%
%\begin{equation}
$\varepsilon_{\rm edge}^{\rm LZ} = \int_{K_+}^{K_{\rm f}} \frac{\ud k}{2\pi} \; \left[ E_{k,+} - E_{k,-} \right]$,
%\end{equation}
%
where $E_{k,+/-}$ are the right/left final edge bands.
Starting the QA evolution from negative $\Delta_{AB}/t_2$, or having $\phi_H=-\frac{\pi}{2}$,
swaps the role of right and left, and of the two Dirac points.
%
%....................
%.................

{\em Adiabatic Floquet Results.} 
We now return to our Floquet QA, Eq.~\eqref{H_gen:eqn}, with a circularly polarized driving. 
We performed an ``adiabatic'' linear turning-on of $\lambda(t)=(t/\tau_{\QA})\lambda_{\rm f}$ for a time $\tau_{\QA}=n_{\QA}\tau$, 
followed by an evolution with constant $\lambda_{\rm f}$ for $\tau_{\rm f} = n_{\rm f}\tau$ (with $\tau=2\pi/\omega$).
We considered both a time-independent $\Delta_{AB}$,
and a $\Delta_{AB}(t)$ switched-off to $0$ during the annealing of $\lambda(t)$ 
(physically possible in the cold-atom realization). 
In principle, $\Delta_{AB}=0$ in graphene, but the initial zig-zag edge states would be pathological  
symmetric/antisymmetric combinations of left-right wavefunctions \cite{Lado_arxiv15}.
Hence, to describe graphene we include a very small $\Delta_{AB}=0^{\pm}$ whose value is not crucial
(only the sign matters); the results turn out to be identical to evolutions in
which $\Delta_{AB}(t)$ is switched-off to $0$.
A generalized adiabatic theorem holds for Floquet systems \cite{Breuer_PLA89,Breuer_ZPD89,Althorpe_CP97,Kitagawa_PRB11,Liu_PRL13}: 
a Floquet state $\ket{\psi_{\alpha} (\lambda(0))}$ evolves remaining close to the instantaneous 
Floquet state $\ket{\psi_{\alpha} (\lambda(t))}$ for sufficiently slow variations of the driving amplitude $\lambda(t)$,
compared to the gaps from neighboring states $\min_{m \in \mathbb{Z}} (|\epsilon_\alpha - \epsilon_\beta + m\hbar\omega|)$.
%
%when the Floquet Brillouin zone (FBZ) distance
%$\min_{m \in \mathbb{Z}} (|\epsilon_\alpha - \epsilon_\beta + m\omega|)$ between 
%two quasienergies is small.
% Since the residual energy is finite for any $\tau_{\QA}$ and
%  we cannot define a ``residual quasienergy'',
%
For $\hbar\omega > W$, the standard Floquet BZ $[-\hbar \omega/2, \hbar \omega/2]$ is
such that the initial Slater determinant $\ket{\Psi(0)}$ coincides with the 
\textit{Floquet ground state} $\ket{\Psi_{\rm FGS}(\lambda(0)=0)}$:
all negative quasi-energies, in $[-\hbar\omega/2,0)$, are occupied, the positive ones are empty.
The only relevant gap for the adiabatic Floquet dynamics \cite{Breuer_PLA89,Breuer_ZPD89} is that at the
Dirac points. 
A slow increase of $\lambda$ will reproduce the QA of the Haldane case, as we verified 
by monitoring the occupations $n_{k,\alpha}$ of the instantaneous single-particle Floquet modes 
$|\phi_{k,\alpha}\rangle$ \cite{SM:note}:
the state $\ket{\Psi(\tau_\QA)}$ after the annealing is ``close'' to $\ket{\Psi_{\rm FGS}(\lambda(\tau_\QA))}$, 
apart from bulk KZ excitations near the Dirac points, and the previously discussed selective excitation of edge states, 
see Fig.~\ref{Graphene1:fig}-${\bf a}$.

%The usual residual energy is not suitable to monitor adiabaticity: when $\Delta_{AB}={\rm constant}$, for instance,
%one can show that $E_{\rm gs}(t)$ is constant in time, while the average energy $\langle \Psi(t)|\hat{H}(t)|\Psi(t)\rangle$ 
%increases, and $E_{\rm res}(t)=\langle \Psi(t)|\hat{H}(t)|\Psi(t)\rangle-E_{\rm gs}$ remains finite for $\tau_{\QA}\to \infty$: 
%such constant piece in $E_{\rm res}$ embodies the deformation of the quasi-energy bands 
%from the unperturbed ones. 
%%Unfortunately, a ``residual quasi-energy'' would not be a well defined physical quantity because of the $2\pi/\omega$ periodicity;
%The occupation $n_{k,\alpha}$ of the instantaneous Floquet modes $|\phi_{k,\alpha}\rangle$ 
%(calculated assuming a constant $\lambda(t)$ over a period) is instead a good indicator of adiabaticity.
%
%.................
%................
%
The dynamics of the Hall current is interesting. 
As customary, in a Laughlin cylinder geometry the total current in the $y$-direction is
given by $\hat{J}_y=\frac{1}{\hbar}\frac{\partial \hat{H}\; }{\partial \kappa_y}\big|_{\kappa_y=0}$, 
where $\kappa_y=\frac{2\pi}{N_y a}\frac{\Phi_L}{\phi_0}$ is related to the flux $\Phi_L$, 
piercing the PBC-cylinder along the $x$-axis, and $\phi_0$ is the flux quantum \cite{Laughlin_PRB81}.
With PBC along $y$, we can write 
$\hat{H}(t) = \sum_{k}^{{\rm BZ}_y} \sum_{ii'} \mathbb{H}_{ii'}^{\phantom \dagger}(k,t) \; \opcdag{i,k} \opc{i',k}$,
where $\mathbb{H}(k,t)$ is the $k$-resolved strip Hamiltonian and 
$\opcdag{i,k}$ creates an electron of momentum $k$ at site $i$ along the zig-zag line 
sketched in Fig.~\ref{Graphene1:fig}-${\bf b'}$.
Hence, bond-resolved $y$-currents are 
$J_{i,i+1}(t)=\langle\Psi(t) | \sum_{k}^{{\rm BZ}_y} \mathbb{J}_{i,i+1}^{\phantom \dagger}(k,t) \; \opcdag{i,k} \opc{i+1,k} | \Psi(t)\rangle$
with $\mathbb{J}=\frac{1}{\hbar}\frac{\partial \mathbb{H}\; }{\partial \kappa_y}\big|_{\kappa_y=0}$.
The circles in Fig.~\ref{Graphene1:fig}-${\bf a'}$ denote the time-average of
$J_{i,i+1}(t)$ during the constant-$\lambda_{\rm f}$ evolution, 
$[J_{i,i+1}]_{\rm av} = \frac{1}{n_{\rm f}\tau} \int_{\tau_{\QA}}^{\tau_{\QA} + n_{\rm f}\tau} \! \ud t \; J_{i,i+1}(t)$.
%
%Different time-averages of $J_y$ flowing over the nn bonds
%$(i,i+1)$, at $t>\tau_{\QA}$, are shown in : 
%the open squares are stroboscopic averages of currents taken at times $n\tau$, averaged over $n$ for 
%$n\ge n_{\rm f}>n_{\QA}$ (to allow for a transient after the QA):
%%
%$[J_{i,i+1}]_{\rm strobo} = \frac{1}{N+1} \sum_{n=n_{\rm f}}^{n_{\rm f}+N} J_{i,i+1}(t=n\tau)$, 
%%
%%(an average which can be extracted by calculating a Floquet-diagonal Green's function), 
%The two estimates differ (see also SM), due to the large time-periodic oscillations shown in the inset: 
%
Currents are concentrated at the edges, but with large periodic oscillations, shown in the inset:
the stroboscopic averages are not representative of the true time-averages \cite{SM:note}.
%
%In particular, we notice that the stroboscopic average current is sizable in the bulk of the sample for the evolution
%with constant $\Delta_{AB}=0.1|t_1|$, while the actual time-averages show that the current is essentially concentrated at the
%edges. 
%When $\Delta_{AB}$ is infinitesimally small (${\bf b'}$) or annealed to $0$ (data not shown, but
%identical to ${\bf b'}$), the bulk contribution is essentially absent even from the stroboscopic averages, 
%but the periodic oscillations remain very large. 
%In any case, for sufficiently long $\tau_{\rm \QA}$,  not only the bulk is on the  insulating over a period, as expected, 
%but $J_y$ goes to zero even on each bulk bond, 
%
Notice that the edge currents have left/right symmetry, while one would naively expect currents only on
the edge which is selectively occupied by the out-of-equilibrium dynamics (the right edge, for Fig.~\ref{Graphene1:fig}). 
This behavior originates from specific symmetries --- previously
noted in equilibrium for the Haldane model at $\phi_H = \pm \pi/2$ \cite{Caio_PRL15,Dalhaus_PRL15} --- 
whereby the GS value of $\hat{J}_y$ is {\it zero everywhere} due to an exact compensation between 
currents due to edge states and edge-current contributions due to {\em bulk} states. 
Since the out-of-equilibrium dynamics brings a lack of current-carrying edge states 
(at left, in Fig.~\ref{Graphene1:fig}-{\bf a}), the corresponding bulk contribution is uncompensated 
and gives rise to a left-flowing current of the same sign and amplitude as that at the right-edge.
This left/right symmetry can be removed by a space inhomogeneity of the perturbation, 
e.g., a laser focused off-center. 
We find here expedient to retain translational invariance along $y$, 
assuming a $y$-independent Gaussian modulation amplitude
$A_0(\x,t)=A_0(t) \; \nep^{-\left(x -x_c\right)^2/2\sigma^2}$, where $x_c$ is the
focus center, and $\sigma$ the beam width. 
All the previous results remain valid for a central focusing, $x_c=L_x/2$, provided $\sigma$ is not too small
($\sigma\gtrsim 0.4 L_x$). 
But the interesting new feature is the ability to control the edge current to flow on either edge of the sample by moving the focus off-center.  
Fig.~\ref{Graphene1:fig}$-{\bf b}$ illustrates the final Floquet quasi-energy bands when the laser
is focused on the right edge ($x_c=L_x$), with $\sigma=0.4 L_x$.
Notice that only the irradiated right-edge states show a $k$-dispersion: unirradiated left-edge states
stay flat and carry no current.
%reducing $\sigma$ for $x_c\neq L_x/2$ breaks 
%the left/right symmetry of the edge currents obtained for uniform irradiation: 
%for $\sigma \sim 0.4 L_x$ 
Non-equilibrium currents flow only at the irradiated edge, see Fig.~\ref{Graphene1:fig}-${\bf b'}$.
%But notice that the edge current is carried by right-edge states in \revision{${\bf a'}$}. 
%
% Per ora ometto ....
%\revision{Results (not shown) for a focusing on the left edge ($x_c=0$) with the same initial condition
%show that while the right-edge states are again selectively populated, they stay flat since they are not irradiated;
%left-edge states now acquire a dispersion, but carry no current since they are not populated during the dynamics: 
%the current flowing on the left side is all due to bulk states.}

%.................
\begin{figure}
\begin{center}
  \includegraphics[width=8cm]{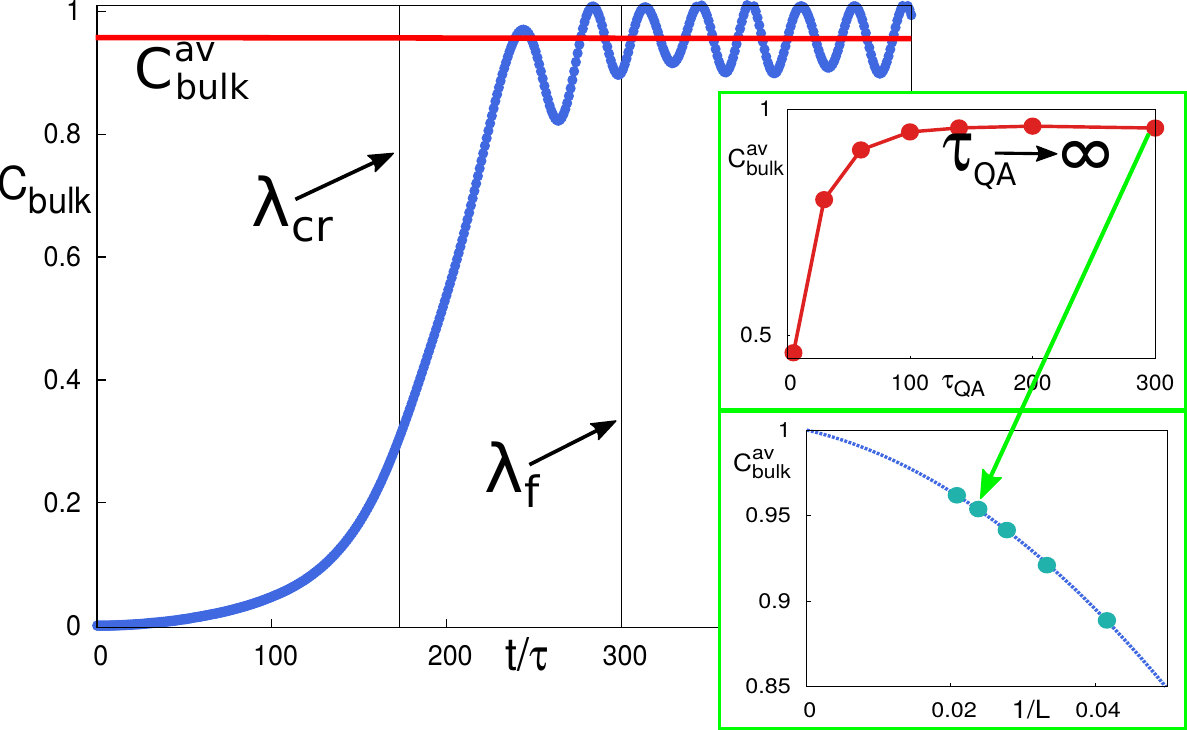}
\end{center}
\setlength{\abovecaptionskip}{0cm}
\caption{(Color online) The bulk average $C_{\rm bulk}(t)$ of the local Chern marker $\calC(\bfr,t)$
for a uniform driving with $\phi = - \pi/2$, $\hbar \omega = 7 |t_1| > W$, $\Delta_{AB}=0.1|t_1|$, 
and $\lambda(t)$ linearly ramped up to $\lambda_{\rm f}=1$ in $\tau_{\QA}=300 \tau$,
followed by a constant-$\lambda_{\rm f}$ evolution for $\tau_{\rm f} = 220 \tau$, with $\tau=2\pi/\omega$.
The topological transition occurs at $\lambda_{\rm cr}\approx 0.57$.  
Here $L=N_x=N_y=48$, and we average on a central square of size $12 \times 12$. 
The horizontal line at $\approx 0.96$ is the time-average $C_{\rm bulk}^{\rm av}$, 
calculated from $t=\tau_{\QA}$ to $t = \tau_{\QA} +\tau_{\rm f}$. 
The upper inset shows the saturation of $C_{\rm bulk}^{\rm av}(L,\tau_{\QA}\to\infty)$ to 
a limiting value that, see lower inset (where we fit points with standard power-law corrections $1+c_1/L+c_2/L^2$), 
goes to $1$ for $L\to\infty$. }
\label{FigChern:fig}
\end{figure}
%.................
%
It is interesting to address the issue of an ``indicator'' of non-trivial topology in a
non-equilibrium translationally-non-invariant setting.
For translational invariant systems (with PBC), it was shown that the usual Chern number $\calC$ is conserved during 
a unitary evolution \cite{Dalessio_NatCom15,Caio_PRL15}: it does not ``signal'' the topological transition. 
However, since for $\hbar\omega>W$ the non-trivial final bulk states are adiabatically populated in a controlled way, 
we would expect to be able to ``see'' the topological transition by looking only at the ``bulk'' of the sample. 
%because of the nearsightedness of electrons in insulators \cite{Kohn_PRL96}.
We find that the local Chern marker $\calC(\bfr)$ introduced in Ref.~\onlinecite{Bianco_PRB11} at equilibrium,
essentially a local measure of the Hall conductivity, 
works also in our non-equilibrium context: it signals if the sample bulk is \textit{locally} 
a topologically non-trivial insulator, $\calC(\bfr)\sim \pm 1$.
$\calC(\bfr)$ can be expressed as a physically appealing commutator of position operators 
\cite{Bianco_PRB11,SM:note}:
\begin{equation} \label{eq:ChernMarker_2}
\calC(\bfr,t) =  - 2\pi i \bra{\bfr} \left[ \hat{x}_{{\calP}(t)}, \hat{y}_{{\calP}(t)} \right] \ket{\bfr} \;.
\end{equation}
Here $\hat{x}_{\calP}={\calP}\hat{x}{\calP}$ and $\hat{y}_{\calP}={\calP}\hat{y}{\calP}$ are position operators 
projected on the occupied states, ${\calP}(t)$ being the projector on the time-evolved Slater determinant $|\Psi(t)\rangle$.
Fig.~\ref{FigChern:fig} illustrates the dynamics of $\calC(\bfr,t)$, averaged over a
central ``bulk'' portion of the sample, $C_{\rm bulk}(t)=N_{\rm bulk}^{-1}\sum_{\bfr\in {\rm bulk}} \calC(\bfr,t)$
as the system evolves from a trivial insulator at $\lambda=0$, 
towards the non-trivial point with $\lambda_{\rm f}=1$.
Upon time-averaging the oscillations after $\tau_\QA$ \cite{SM:note}, we obtain a quantity
$C_{\rm bulk}^{\rm av}(L,\tau_{\QA})=\frac{1}{n_{\rm f} \tau} \int_{\tau_{\QA}}^{\tau_{\QA} +n_{\rm f} \tau} \! \! \ud t \; C_{\rm bulk}(t)$
approaching the correct integer value 1 as $\tau_\QA\to \infty$ and $L\to \infty$ 
(see insets in Fig.~\ref{FigChern:fig}).
%$\lim_{L\to \infty} \lim_{\tau_\QA\to \infty} C_{\rm bulk}^{\rm av}(L,\tau_{\QA}) = 1$.
%
%Notice that at all edges (including those imposed by the PBC along $y$) the CM deviates strongly from its bulk behaviour, 
%in such a way that $\int_V \!\ud \bfr \; \calC(\bfr) = 0$ on any finite sample \cite{Bianco_PRB11}.
%Numerically, we find that this is so even when PBC are used in the $y$-direction.

\begin{figure}
\begin{center}
  \includegraphics[width=8cm]{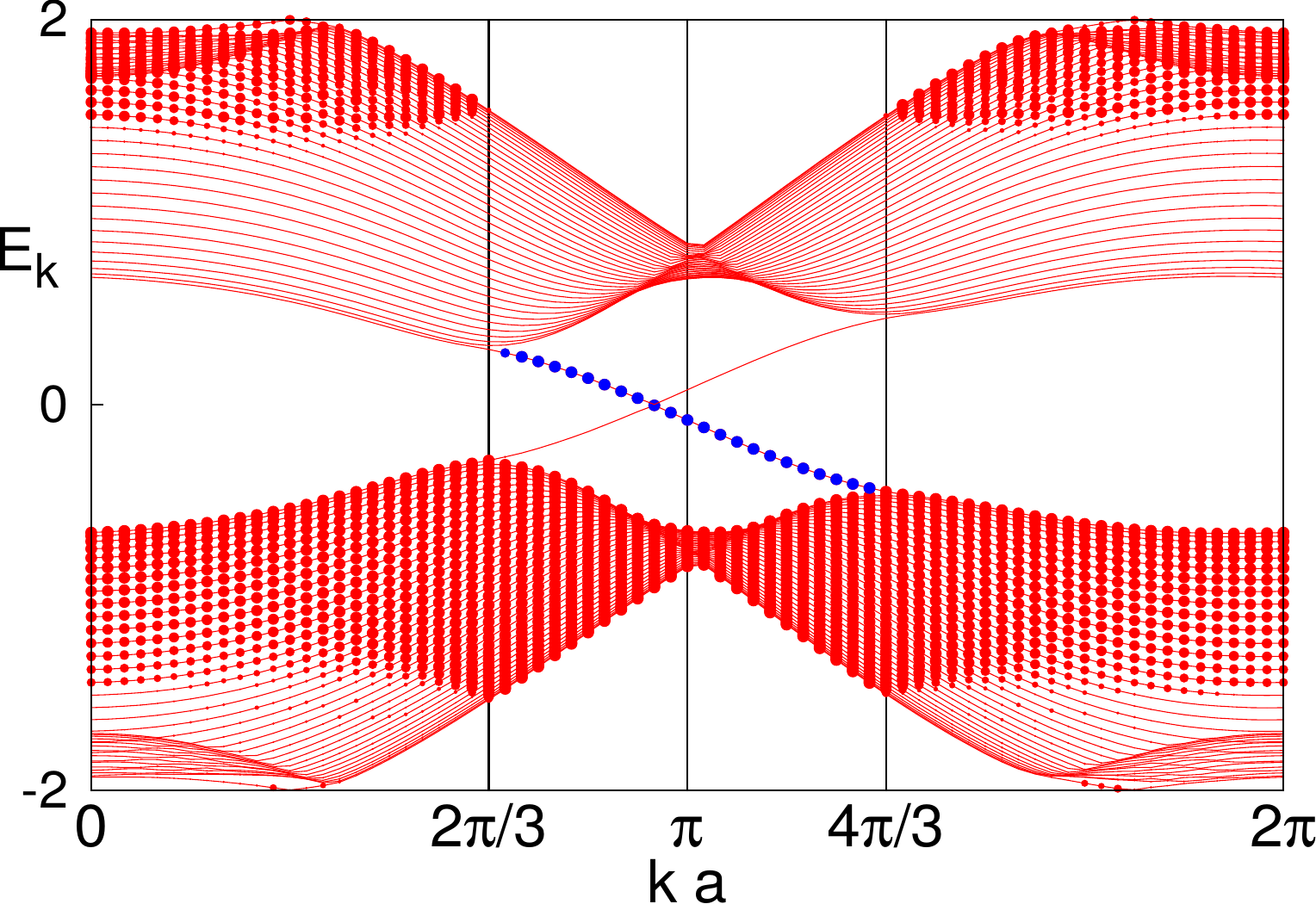}\\
\end{center}
\setlength{\abovecaptionskip}{0cm}
\caption{(Color online) 
Final Floquet quasi-energies $E_{k,\alpha}$ and occupations $n_{k,\alpha}$ (dot size proportional to $n_{k,\alpha}$) 
for $\phi = - \pi/2$, $\hbar \omega = 4|t_1| < W$, $\Delta_{AB}=0.1 |t_1|$ (constant), 
and $\lambda(t)$ linearly ramped up to $\lambda_{\rm f}=1$ in $\tau_{\QA}=300 \tau$.
}
\label{Graphene1bis:fig}
\end{figure}

%{\em Results: resonant case $\hbar\omega<W$.}
%
When $\hbar \omega$ is {\em smaller} than the bandwidth $W$, 
intra-band resonances between valence states at $\epsilon_{\alpha}$ and conduction ones
at $\epsilon_{\beta}\approx \epsilon_{\alpha}+\hbar\omega$ change the physics completely, 
breaking the Floquet adiabatic picture. 
The whole crux is, in essence, that the starting point at $\lambda(0)=0$ 
--- the usual Slater determinant $\ket{\Psi(0)}$ --- does not coincide with
the Floquet GS, $\ket{\Psi(0)}\neq \ket{\Psi_{\rm FGS}(\lambda(0)=0)}$, for
any choice of the Floquet BZ: upon folding the original $[-W/2,W/2]$ spectrum in e.g., $[-\hbar\omega/2,\hbar\omega/2]$, 
we see that the lower Floquet quasi-energy band is only partially filled, with empty conduction-band-originated
states intermixed with filled valence-band ones.
These filled-empty pairs with small Floquet gaps \cite{Breuer_PLA89,Breuer_ZPD89} 
$\min_{m \in \mathbb{Z}} (|\epsilon_\alpha - \epsilon_\beta + m\hbar\omega|)$ lead to a proliferation 
of {\em bulk} LZ events as $\lambda(t)$ is ramped up, with a complex redistribution of electronic
occupations of the states: the final Floquet GS $\ket{\Psi_{\rm FGS}(\lambda(\tau_{\QA})}$ 
is {\em never} reached, even for $\tau_{\QA}\to \infty$. 
The analysis of these issues will be the subject of a future work \cite{Privitera_unpub}. 
%because large spectral gaps at the
%BZ boundaries $\pm \hbar\omega/2$ prevent the correct ground-state population of the lower band.
%
Fig.~\ref{Graphene1bis:fig} shows the final Floquet quasi-energy bands, with the corresponding population
indicated by a variable-size dot, for $\hbar\omega=4|t_1|<W$ and $\lambda(t)$ linearly ramped up to $\lambda_{\rm f}=1$
in $\tau_{\QA}=300 \tau$.
The final state reached is a non-equilibrium metal, rather than an insulator.
%When $\hbar \omega < W$ however, intra-band resonances prevent any smooth adiabatic driving,
%and the FGS has nothing to do with the out-of-equilibrium dynamics of the system.  
%Indeed, as apparent from Fig.~\ref{Graphene1:fig} ${\bf c}$, the final state reached after the annealing can be 
%regarded as a non-equilibrium metal, rather than an insulator. 
This aspect is important in devising pump-probe photoemission experiments in graphene \cite{Sentef_Naturecomm15}.
%
%.................
%................

{\em Conclusions.}  
We found a non-equilibrium mechanism which selectively populates edge states
when performing an adiabatic switching-on of a periodic perturbation towards a topologically non-trivial
insulating phase. 
It is different from the ``topological blocking'' of Ref.~\cite{Kells_PRB14}, 
which works with PBCs and when driving systems with symmetry-protected subspaces 
{\it from} the topologically non-trivial phase {\it to} the trivial one.
The mechanism we illustrated requires edge states (hence OBC) whose electronic occupation
is unable to follow instantaneous equilibrium as they become topologically non-trivial and cross the bulk gap.
It is general enough, and is at the root of the deviations from KZ scaling in 1d topological transitions, 
as seen in \cite{Bermudez_NJP10,Bermudez_PRL11}.
In the present 2d context, it adds flexibility to the control of the edge currents flowing at
the boundaries of the sample, including the ability to have currents flowing only at one edge, by appropriate
focusing of the ac field. 
Finally, we have shown that for $\hbar\omega<W$ intra-band resonances
ruin the adiabatic picture and the resulting state is a non-equilibrium metal.
Our findings should be amenable to experimental tests both with ultra-cold atoms in optical lattices \cite{Jotzu_Nature14}, 
and with laser irradiated electronic systems.

We acknowledge discussions with I. Carusotto, A. Dutta, R. Fazio, A. Russomanno, A. Silva and E. Tosatti.
Research was supported by MIUR PRIN-2010LLKJBX-001, 
and by EU ERC MODPHYSFRICT.

%\bibliography{QIsing,ThesisBiblio,LNBiblio}

\end{document}